\documentclass[twocolumn]{aastex631}
\usepackage{amsmath}

\begin{document}

\title{Weak Lensing Approximation of Wave-optics Effects from General Symmetric Lens Profiles}

\author[0000-0002-0196-9169]{Zhao-Feng Wu}
\email{Email: wu2177@purdue.edu}
\affiliation{Department of Physics and Astronomy, Purdue University, 525 Northwestern Avenue, West Lafayette, IN 47907, USA}
\author[0000-0002-3887-7137]{Otto A. Hannuksela}
\affiliation{Department of Physics, The Chinese University of Hong Kong, Shatin, N.T., Hong Kong}
\author[0000-0001-8322-5405]{Martin Hendry}
\affiliation{SUPA, School of Physics and Astronomy, University of Glasgow, Glasgow G12 8QQ, UK}
\author[0000-0002-1828-3702]{Quynh Lan Nguyen}
\affiliation{Phenikaa Institute for Advanced Study, Phenikaa University, Hanoi 12116, Vietnam}


\begin{abstract}
\noindent Gravitational lensing of electromagnetic (EM) waves has yielded many profound discoveries across fundamental physics, astronomy, astrophysics, and cosmology. Similar to EM waves, gravitational waves (GWs) can also be lensed. When their wavelength is comparable to the characteristic scale of the lens, wave-optics (WO) effects manifest as frequency-dependent modulations in the GW waveform. These WO features encode valuable information about the lensing system but are challenging to model, especially in the weak lensing regime, which has a larger optical depth than strong lensing. We present a novel and efficient framework to accurately approximate WO effects induced by general symmetric lens profiles. Our method is validated against numerical calculations and recovers the expected asymptotic behavior in both high- and low-frequency limits. Accurate and efficient modeling of WO effects in the weak lensing regime will enable improved lens reconstruction, delensing of standard sirens, and provide a unique probe to the properties of low-mass halos with minimal baryonic content, offering new insights into the nature of dark matter.
\end{abstract}

\keywords{Gravitational wave astronomy (675); Gravitational lensing (670); Weak gravitational lensing (1797)}


\section{Introduction}
When electromagnetic (EM) waves propagate near massive objects over cosmological distances, they experience gravitational lensing. 
Over the past several decades, gravitational lensing has become a powerful tool in the EM domain, driving advances across fundamental physics, astronomy, astrophysics, and cosmology \citep{2010CQGra..27w3001B}.
According to the theory of general relativity, gravitational waves (GWs), similarly to EM waves, are also subject to gravitational lensing \citep{Ohanian1974OnRadiation, Deguchi1986DiffractionMass, Wang1996GravitationalBinaries, Nakamura1998GravitationalLens, 2003ApJ...595.1039T}. 
However, because the GWs observable by current and most future planned instruments have significantly longer wavelengths than typical observable EM waves, they can experience diffraction as they travel through astrophysical structures of sizes comparable to their wavelengths. 
The low frequencies and phase coherence of sources observable with the Laser Interferometer Space Antenna (LISA; \citealp{2017arXiv170200786A}) and Pulsar Timing Arrays (PTA; \citealp{1990ApJ...361..300F}) make them particularly ideal for probing lensing by large-scale structures in the wave-optics (WO) regime \citep{2003ApJ...595.1039T,2023Univ....9..200G,2025RSPTA.38340134S} 

WO effects imprint frequency-dependent amplitude and phase modulations on GW signals. Detecting these lensing-induced features enables direct measurement of lens parameters---such as the redshifted lens mass, $M_{\mathrm{Lz}} = (1 + z_{\mathrm{L}}) M_{\mathrm{L}}$, where $z_{\mathrm{L}}$ is the lens redshift, and the source position in the source plane---for simple point-mass and SIS lens models \citep{2023PhRvD.108l3543C,2023PhRvD.107d3029C,2023PhRvD.108j3532S}.
In contrast, identifying lensing signatures in the geometric-optics (GO) regime, corresponding to the high-frequency limit, relies on associating and interpreting multiple images of the same event~\citep{Haris2018IdentifyingMergers,Hannuksela2019SearchEvents, Hannuksela2020LocalizingLensing, 2023PhRvD.107f3023C, Janquart2021AEvents, 2025ApJ...980..258B, Wempe2024OnLensing,2025PhRvD.111f3012C} or so-called Type-II image distortions~\citep{Dai2017OnWaves,2020PhRvD.102l4048E,2021PhRvD.103j4055W,Janquart2021OnModes}.

However, WO features require lenses to have structure within a restricted mass/length range, determined by the frequency band of the observed signal. 
If the lens is too massive/sizable, the system enters the GO regime. 
If, on the other hand, the mass/size is too low, the GW wavelength exceeds the characteristic scale of the lens, and the wave remains unperturbed \citep{2021MNRAS.503.3326C, Mishra2024ExploringDetection,Yeung2023DetectabilityWaves, Mishra2024ExploringDetection, 2023PhRvD.108d3527T}. 
Indeed, GWs typically only couple with astrophysical structures with sizes comparable to their wavelength through WO effects. 
This constraint significantly reduces the probability of detecting WO features, particularly for strongly lensed signals where the most probable lenses are massive galaxies \citep{2024SSRv..220...87S}. 

WO features can also be searched for in weakly lensed signals, which do not require close alignment between the source, lens, and observer, and therefore have a higher probability of occurrence. 
It has been estimated that LISA may detect WO effects at approximately ten times the strong-lensing impact parameter, corresponding to about 1\% of massive black hole binaries \citep{2022MNRAS.512....1G, 2023PhRvD.108l3543C}. 
For PTAs, the first detected single source may lie at high redshift \citep{2024ApJ...972...29W}, offering a potential opportunity to observe WO effects, although detection will largely depend on improvements in signal-to-noise ratio \citep{2025PhRvL.134m1001J}. 
The trade-off between strong and weak lensing is thus between rare, dramatic signatures and frequent, subtle modulations.

A further complication is that accurately computing lensed waveforms in the WO regime is challenging, particularly in the weak lensing limit. 
A closed-form analytical expression exists only for the simplest point-mass lens \citep{PhysRevD.9.2207}, and series expansions have been developed for the singular isothermal sphere (SIS) lens model \citep{2006JCAP...01..023M}. 
However, even these expressions are computationally expensive to evaluate, especially at high frequencies or large impact parameters \citep{2023PhRvD.107d3029C}. 

More generally, predicting WO features for arbitrary lens models requires evaluating integrals of rapidly oscillating functions over the lens plane \citep[see, e.g.,][]{DIego2019ObservationalFrequencies,2021MNRAS.503.3326C,Yeung2023DetectabilityWaves,Yeung2024Wolensing:Effects,2025PhRvD.111j3539V}, a process that is notoriously difficult and poses significant challenges to maintaining precision, particularly in the weak lensing regime.

Therefore, efficient approximations are necessary to accelerate lensing calculations, particularly in the context of population studies. 
Several methods exist for the high- and low-frequency limits \citep[e.g.,][]{2004A&A...423..787T,PhysRevD.104.063001,2023PhRvD.108d3527T,2025PhRvD.111h3541Y}; however, these do not target the intermediate dimensionless frequency range where WO features are most prominent. 
\citet{2023PhRvD.108j3532S} proposed a weak lensing approximation formulated in the time domain of the amplification factor. However, their method requires dedicated regularization and still involves the evaluation of numerical integrals.

In this paper, we present a novel and efficient method to approximate WO effects induced by general symmetric lenses in the weak lensing regime. Section~\ref{Lesing_ofGW} reviews the general formalism of GW lensing. 
In Section~\ref{weak_len_app}, we develop the weak lensing approximation, beginning with the SIS lens in Section~\ref{Sec_SIS}, and extending to general lenses in Section~\ref{Sec_GL}. In Section~\ref{Sec_NFW}, we validate the approximation using the Navarro-Frenk-White (NFW) lens model. Section~\ref{freq_asym} demonstrates the consistency of our method with known approximations in the high- (Section~\ref{high_w}) and low-frequency (Section~\ref{low_w}) limits, while Section~\ref{discussion} discusses the advantages and implications of the method. We conclude with potential applications and directions for future work in Section~\ref{conclu}.

\section{Lensing of Gravitational Wave} \label{Lesing_ofGW}
For a comprehensive overview of gravitational lensing theory, we refer the reader to \citet{2002glml.book.....M} and \citet{1992grle.book.....S}. 
In the frequency domain, the effect of lensing is characterized by a multiplicative factor $F(f)$, known as the amplification factor \citep{2003ApJ...595.1039T}:
\begin{equation}
    F(f) \equiv \frac{\tilde{h}(f)}{\tilde{h}_0(f)},
\end{equation}
where $\tilde{h}_0(f)$ and $\tilde{h}(f)$ are the Fourier transforms of the unlensed and lensed strain amplitudes, respectively. 
The simplicity of this relation arises from the fact that, apart from their dependence on frequency, the source parameters defining the unlensed signal and the lens parameters defining the amplification factor are uncorrelated. 
The amplification factor depends solely on the mass density profile of the lens and the source position with respect to it.

The frequency-domain amplification factor in dimensionless form is given by\footnote{Here we adopt the Physics convention for the Fourier transform.}
\begin{equation} \label{F(w)}
F(\omega, \boldsymbol{y}) = \frac{\omega}{2 \pi i} \int \mathrm{d}^2 \boldsymbol{x} \, \exp \left( i \omega \phi(\boldsymbol{x}, \boldsymbol{y}) \right),
\end{equation}
where $\omega$ is the dimensionless frequency, $\boldsymbol{x}$ and $\boldsymbol{y}$ are the dimensionless coordinates in the lens plane and source plane, respectively. Readers are referred to \citet{2003ApJ...595.1039T} for detailed derivations. The integration is performed over the lens plane, with $\boldsymbol{x}$ rescaled by a characteristic scale $\xi_0$ of the lens. The source position $\boldsymbol{y}$ is rescaled by $\eta_0 \equiv D_S \xi_0 / D_L$, where $D_S$ and $D_L$ are the angular diameter distances to the source and lens, respectively. The dimensionless frequency $\omega$ is defined as~\citep{Dai2017OnWaves,2021PhRvD.103f4047E}
\begin{equation}
\omega \equiv \frac{8 \pi G M_{\mathrm{Lz}} f}{c^3},
\end{equation}
in terms of the redshifted effective lens mass
\begin{equation} \label{M_Lz}
M_{\mathrm{Lz}} \equiv \frac{\xi_0^2 c^2}{2 G d_{\mathrm{eff}}},
\end{equation}
where $d_{\mathrm{eff}} \equiv \frac{D_L D_{\mathrm{LS}}}{(1 + z_L) D_S}$ depends on the angular diameter distance between the lens and the source, $D_{\mathrm{LS}}$. For a point-mass lens, $M_{\mathrm{Lz}}$ coincides with the total lens mass (i.e., setting $\xi_0$ equal to the Einstein radius), but this is not generally true for extended lenses.\footnote{See, e.g., \citet{2023PhRvD.108j3532S} for the case of lensing by a singular isothermal sphere.}

The integral depends on the Fermat potential:
\begin{equation}
\phi(\boldsymbol{x}, \boldsymbol{y}) = \frac{1}{2} |\boldsymbol{x} - \boldsymbol{y}|^2 - \psi(\boldsymbol{x}) - \phi_m(\boldsymbol{y}),
\end{equation}
which is the dimensionless version of the time delay. Here, $\psi(\boldsymbol{x})$ is the lensing potential, determined by the projected matter distribution on the lens plane, and its gradient yields the deflection angle. Specifically, it satisfies
\begin{equation}
\nabla_{\boldsymbol{x}}^2 \psi(\boldsymbol{x}) = \frac{2 \Sigma\left(\xi_0 \boldsymbol{x}\right)}{\Sigma_{\mathrm{cr}}},
\end{equation}
where $\nabla_{\boldsymbol{x}}^2$ is the two-dimensional Laplacian, $\Sigma\left(\xi_0 \boldsymbol{x}\right)$ is the projected surface mass density, and $\Sigma_{\mathrm{cr}} \equiv \left(4 \pi G (1 + z_L) d_{\mathrm{eff}}\right)^{-1}$ is the critical surface density. The term $\phi_m(\boldsymbol{y})$ is a constant with respect to the integration, and is conventionally chosen such that the global minimum of the Fermat potential is set to zero.

An important limiting case of WO lensing is the GO regime, corresponding to the $\omega \to \infty$ limit of the diffraction integral (Equation~\ref{F(w)}):
\begin{equation}
F(\omega) = \sum_J \sqrt{|\mu_J|} \, e^{i \omega \phi_J - i \text{sgn}(\omega) \pi n_J}.
\end{equation}
Here, the index $J$ labels the GO images, located at stationary points $\boldsymbol{x}_J$ of the Fermat potential, satisfying $\phi_{,i}(\boldsymbol{x}_J, \boldsymbol{y}) = 0$, where a comma subscript denotes differentiation with respect to lens-plane coordinates. The magnification is given by
\begin{equation}
\mu_J^{-1} \equiv \det \left( \phi_{,ij}(\boldsymbol{x}_J) \right),
\end{equation}
and the time delay is $\phi_J \equiv \phi(\boldsymbol{x}_J, \boldsymbol{y})$, both evaluated at the image positions. The Morse phase $n_J$ is 0, $\pi/2$, or $\pi$ depending on whether $\boldsymbol{x}_J$ corresponds to a minimum, saddle point, or maximum of $\phi$, respectively. In the single-image regime, the GO limit reduces to a simple rescaling of the waveform, $F(\omega) = \sqrt{\mu}$.

For any symmetric lens, the two-dimensional amplification factor integral can be reduced to a one-dimensional form \citep{2003ApJ...595.1039T}:
\begin{equation} \label{general_sym}
\begin{aligned}
F(\omega, y) = & -i \omega \, e^{i \omega y^2 / 2} \int_0^{\infty} x J_0(\omega x y) \\
& \times \exp \left[ i \omega \left( \frac{1}{2} x^2 - \psi(x) - \phi_m(y) \right) \right] \, \mathrm{d}x,
\end{aligned}
\end{equation}
where $\omega$, $\psi(x)$, and $\phi_m(y)$ are as previously defined, and $y$ is the magnitude of the rescaled source position. In this paper, we focus on symmetric lenses.

\section{Weak Lensing Approximation} \label{weak_len_app}
For a general lensing potential $\psi(x)$, there is no closed-form expression for the 
amplification factor $F(\omega, y)$, and numerical integration is typically required. 
For relatively simple lens models, such as the SIS, analytical solutions exist; however, the resulting expressions for $F(\omega, y)$ remain complicated and require careful implementation at large $y$ or $\omega$ \citep{2006JCAP...01..023M, 2023PhRvD.108j3532S}. 
In practice, the probability that a GW signal undergoes strong lensing is lower than it is for weak lensing \citep{Ng2017PreciseHoles, Li2018GravitationalPerspective, 2018MNRAS.480.3842O, 2022ApJ...929....9X, Wierda2021BeyondLensing, More2022ImprovedEvents, Phurailatpam2024LerSimulator}. 
Therefore, in this work, we try to develop a weak lensing approximation to efficiently estimate the amplification factor.

In the GO regime, weak lensing corresponds to sources located far from the lens, or equivalently $y \gg 1$. In this limit, the condition $y > y_{\mathrm{cr}}$, where $y_{\mathrm{cr}}$ is the critical impact parameter for multiple images, is automatically satisfied, placing the system in the single-image regime.\footnote{For typical lensing profiles under standard definitions, $y \sim 10$ is already considered large compared to $y_{\mathrm{cr}}$. We adopt this convention for $y \gg 1$ throughout this work.}
In the WO regime, an additional condition must be satisfied: $\omega y^2 / 2 \gg 1$\footnote{In this work, values of order several tens are considered sufficiently large.}, as we demonstrate below. This condition is automatically met as $\omega \to \infty$, recovering the GO limit.

We first derive the weak lensing approximation for the SIS model. Motivated by this derivation, we then propose a prescription of approximation for general symmetric lens profiles. A rigorous mathematical proof is beyond the scope of this work; instead, we validate the approximation by comparing it with numerical results for the NFW lens model, an important lensing potential for which no accurate analytical expression is currently available.

\subsection{Singular Isothermal Sphere} \label{Sec_SIS}
The singular isothermal sphere, or SIS, profile is one of the simplest and most widely used models for describing the mass distribution of galaxies, characterized by a flat rotation curve. This behavior arises from modeling the galaxy as an extended system of luminous matter embedded within a dark matter halo. Following the approach of \citet{2006JCAP...01..023M}, we first derive the analytical expression for the amplification factor $F(\omega, y)$ for the SIS model and then develop the weak lensing approximation.

The full density profile of the SIS model is given by \citet{1987gady.book.....B}: 
\begin{equation}
\rho_{\mathrm{SIS}}(r) = \frac{\sigma^2}{2 \pi G r^2},
\end{equation}
where $\sigma$ is the one-dimensional velocity dispersion of stars in the galaxy. The normalization of the SIS profile is related to $\sigma$ by \citet{2003ApJ...595.1039T}:
\begin{equation}
\xi_0 = \frac{4 \pi \sigma^2}{c^2} \frac{D_{\mathrm{L}} D_{\mathrm{LS}}}{D_{\mathrm{S}}},
\end{equation}
where $D_{\mathrm{L}}$, $D_{\mathrm{LS}}$, and $D_{\mathrm{S}}$ are defined as before. 
The corresponding lensing potential is $\psi_\text{SIS}(x) = x$, with $\phi_\text{m,SIS}(y) = -(y + 1/2)$, then the general expression for the amplification factor of an SIS model is
\begin{equation} \label{eq:F_sis}
\begin{aligned}
F_\text{SIS}(\omega, y) = & -i \omega \, e^{i \omega y^2 / 2} \int_0^{\infty} x J_0(\omega x y) \\
 & \times \exp \left[ i \omega \left( \frac{1}{2} x^2 - \psi_\text{SIS}(x) + \phi_\text{m,SIS}(y) \right) \right] \, \mathrm{d}x.
\end{aligned}
\end{equation}
Expanding the exponential term involving the lensing potential yields
\begin{equation} \label{Eq9}
\begin{aligned}
F_\text{SIS}(& \omega, y) =  C_\text{SIS}(\omega,y) \\
& \times \sum_{n=0}^{\infty} \frac{(-i \omega)^n}{n!} \int_0^{\infty} x^{1+n} J_0(\omega x y) \exp\left( \frac{1}{2} i \omega x^2 \right) \, \mathrm{d}x\,,
\end{aligned}
\end{equation}
where 
\begin{equation}
    C_\text{SIS}(\omega,y)=-i \omega \, e^{\frac{i}{2} \omega y^2 - i \omega \phi_m(y)}\,. 
\end{equation}
Using the integral representation of the confluent hypergeometric function of the first kind \citep{1954tit..book.....B},
\begin{equation} \label{int_rep}
{ }_1F_1(a, b; -z) = \frac{z^{\frac{1}{2} - \frac{1}{2}b}}{\Gamma(a)} \int_0^{\infty} t^{a - b/2 - 1/2} J_{b-1}(2 \sqrt{z t}) \exp(-t) \, \mathrm{d}t,
\end{equation}
we make the substitutions $z = i \omega y^2/2$, $t = -i \omega x^2/2$, $a = 1 + n/2$, and $b = 1$. Substituting the integral representation of the confluent hypergeometric function (Equation~\ref{int_rep}) into the amplification factor (Equation~\ref{Eq9}), we obtain the result of \citet{2006JCAP...01..023M}:
\begin{equation} \label{SIS_full}
\begin{aligned}
&F(\omega, y) =  \, e^{\frac{i}{2} \omega \left( y^2 - 2 \phi_m(y) \right)} \sum_{n=0}^{\infty} \frac{\Gamma\left(1 + \frac{n}{2}\right)}{n!} \left( -i 2 \omega \right)^{n/2} \\
& \quad \times { }_1F_1\left(1 + \frac{n}{2}, 1; -\frac{i}{2} \omega y^2\right) \\
&=  \, e^{-i \omega \phi_m(y)} \sum_{n=0}^{\infty} \frac{\Gamma\left(1 + \frac{n}{2}\right)}{n!} \left( -i 2 \omega \right)^{n/2} { }_1F_1\left( -\frac{n}{2}, 1; \frac{i}{2} \omega y^2 \right).
\end{aligned}
\end{equation}
Note that this above (Equation~\ref{SIS_full}) provides the full solution for the SIS model, valid in both the strong and weak lensing regimes. Motivated by the argument of the confluent hypergeometric function, we propose an additional condition for the weak lensing approximation in the WO regime, requiring $\omega y^2 / 2 \gg 1$\footnote{The physical description of this condition will be discussed later in Section~\ref{Sec_GL}.}. Under this condition, the asymptotic expansion of the confluent hypergeometric function can be used to approximate $F(\omega, y)$ at large distances, which is given by \citet{1970hmfw.book.....A}: 
\begin{equation} \label{Asypm}
\begin{aligned}
& \frac{{ }_1 F_1(a, b; z)}{\Gamma(b)}=  \\
& \frac{e^{i \pi a} z^{-a}}{\Gamma(b-a)}\left\{\sum_{n=0}^{R-1} \frac{(a)_n(1+a-b)_n}{n!}(-z)^{-n}+O\left(|z|^{-R}\right)\right\} \\
& +\frac{e^z z^{a-b}}{\Gamma(a)}\left\{\sum_{n=0}^{s-1} \frac{(b-a)_n(1-a)_n}{n!} z^{-n}+O\left(|z|^{-s}\right)\right\}
\end{aligned}
\end{equation}
valid for $|z| \gg 1$ and $-\frac{1}{2} \pi < \arg z < \frac{3}{2} \pi$. In our case, $|z| = \omega y^2 / 2 \gg 1$ and $\arg z = \pi/2$, satisfying these conditions. Substituting $a = -n/2$, $b = 1$, and taking the limits $R \to \infty$ and $s \to \infty$, the confluent hypergeometric function (Equation~\ref{Asypm}) becomes
\begin{equation} \label{WL_expad}
\begin{aligned}
    & { }_1 F_1  \left(-\frac{n}{2}, 1 ;\frac{i}{2} \omega  y^2\right) \\
    & = \frac{e^{ -i \pi \frac{n}{2}} (\frac{i}{2} \omega  y^2)^{\frac{n}{2}}}{\Gamma(1+\frac{n}{2})}\left( \sum_{k=0}^{\infty} \frac{\left((-\frac{n}{2})_k\right)^2}{k!}(-\frac{i}{2} \omega  y^2)^{-k}\right) \\
    & + \frac{e^{\frac{i}{2} \omega  y^2} {(\frac{i}{2} \omega  y^2)}^{-\frac{n}{2} - 1}}{\Gamma(-\frac{n}{2})} \left( \sum_{k=0}^{\infty} \frac{\left((1+\frac{n}{2})_k\right)^2}{k!}(\frac{i}{2} \omega  y^2)^{-k}\right).
\end{aligned}
\end{equation}

The asymptotic expansion of the confluent hypergeometric function yields two distinct contributions. When the prefactors are collected and the two parts are summed separately in the full solution for the SIS model (Equation~\ref{SIS_full}), the resulting terms can be identified as the GO limit and the WO effect, denoted by $T_\mathrm{GO}$ and $T_\mathrm{WO}$, respectively. We treat these contributions separately to highlight their physical distinction. The first term, associated with the GO limit, is
\begin{equation}\label{eq:TGO}
\begin{aligned}
    T_\text{GO}&= e^{-i \omega \phi_m(y)} \sum_{n=0}^{\infty} \frac{\Gamma\left(1+\frac{n}{2}\right)}{n!} \left({- i2\omega }\right)^{n / 2}\frac{(-\frac{i}{2} \omega  y^2)^{\frac{n}{2}}}{\Gamma(1+\frac{n}{2})} \\
    & \times \left( \sum_{k=0}^{\infty} \frac{\left((-\frac{n}{2})_k\right)^2}{k!}(-\frac{i}{2} \omega  y^2)^{-k}\right)\\
    & = e^{-i \omega \phi_m(y)} \sum_{n=0}^{\infty} \frac{(-i \omega  y)^{n}}{n!} \left( 1 + \left(\frac{n}{2}\right)^2 (-\frac{i}{2} \omega  y^2)^{-1} \right. \\
    & + \left. \left(\frac{n}{2}\right)^2 \left(\frac{n}{2} - 1\right)^2 (-\frac{i}{2} \omega  y^2)^{-2} +  \mathcal{O}\left((\omega  y^2)^{-3}\right) \right) \\
    & = e^{-i \omega \phi_m(y)} \left( C_0 + C_1 + C_2 + \ldots \right)
\end{aligned}
\end{equation}
where $C_0 = \sum_{n=0}^{\infty} \frac{(-i \omega  y)^{n}}{n!} = e^{-i \omega y}$. For the next order, 
\begin{equation}
\begin{aligned}
     & C_1 =\sum_{n=0}^{\infty} \frac{(-i \omega  y)^{n}}{n!} \frac{n^2}{2y} (-i \omega y)^{-1} \\
     & = \sum_{n=1}^{\infty} \frac{(-i \omega  y)^{n-1}}{(n-1)!} \left( \frac{n}{2y} \right) \\
     & = \sum_{n=1}^{\infty} \frac{(-i \omega  y)^{n-1}}{(n-1)!} \left( \frac{n-1}{2y} + \frac{1}{2y} \right) \\
     & = \sum_{n=2}^{\infty} \frac{(-i \omega  y)^{n-1}}{(n-2)!} \frac{1}{2y} + \sum_{n=1}^{\infty} \frac{(-i \omega  y)^{n-1}}{(n-1)!} \frac{1}{2y} \\
     & = \frac{(-i \omega  y)e^{-i \omega y}}{2y} + \frac{e^{-i \omega y}}{2y} = e^{-i \omega y}\left(S_{10} + S_{11}\right).
\end{aligned}
\end{equation}
Similarly, the term of second order in $(-\frac{i}{2} \omega  y^2)^{-1}$ can be simplified accordingly and yields 
\begin{equation}
\begin{aligned}
    & C_2 = \sum_{n=0}^{\infty} \frac{(-i \omega  y)^{n}}{n!} \frac{n^2(n-2)^2}{4y^2} (-i \omega  y)^{-2} \\
    & = \frac{e^{-i \omega y}}{8y^2}\left[ (-i \omega y)^2 - 2(-i \omega y) - 1 + (-i \omega y)^{-1}\right] \\
    & = e^{-i \omega y}\left(S_{20} + S_{21} + S_{22} + S_{23}\right)
\end{aligned}
\end{equation}
Combining all the terms up to the second order in $(-\frac{i}{2} \omega  y^2)^{-1}$, we have
\begin{equation} \label{sq_bracket}
\begin{aligned}
    & \sum_{n=0}^{\infty} \frac{(-i \omega  y)^{n}}{n!} \left[ 1 + \frac{n^2}{4} (-\frac{i}{2} \omega  y^2)^{-1} \right. \\
    & \qquad \qquad \quad \left. + \frac{n^2(n-2)^2}{16} (-\frac{i}{2} \omega  y^2)^{-2} +  \mathcal{O}\left((-\frac{i}{2} \omega  y^2)^{-3}\right)  \right] \\
    & = e^{-i \omega y}\left( 1 + S_{10} + S_{11} + S_{20} + S_{21} + S_{22} + S_{23} + \ldots \right)\\
    & = e^{-i \omega y}\left[ 1 + \frac{-i\omega}{2} + \frac{1}{2y}  + \frac{(-i\omega)^2}{8} \right.\\
    & \qquad \qquad \quad \left. - \frac{-i\omega}{4y} - \frac{1}{8y^2} + \frac{1}{-i\omega 8 y^3} + \ldots \right] .
\end{aligned}
\end{equation}
The terms enclosed in the square brackets of Equation~\ref{sq_bracket} can be directly identified with the series expansion of the following expression:
\begin{equation}
\begin{aligned}
    & e^{- i \omega/2}\left( \sqrt{1 + \frac{1}{y}} \right) \left( 1+\frac{i}{8 \omega y^3} + \frac{i}{\omega}\mathcal{O}(y^{-5})\right) \\  
    & \sim \left(1 + (\frac{- i \omega}{2}) + \frac{(- i \omega)^2}{8} + \mathcal{O}(\omega^{3})\right)\left( 1 + \frac{1}{2y} \right. \\
    & \left. \;\; \;\; - \frac{1}{8y^2} + \mathcal{O}(y^{-3}) \right)\left( 1+\frac{i}{8 \omega y^3} + \frac{i}{\omega}\mathcal{O}(y^{-5})\right) \\
    & = 1 + \frac{-i\omega}{2} + \frac{(-i\omega)^2}{8} + \frac{1}{2y} - \frac{-i\omega}{4y} - \frac{1}{8y^2} + \ldots
\end{aligned}
\end{equation}
Combining all terms in $T_\text{GO}$ (Equation~\ref{eq:TGO}), and noting that $\phi_m(y) = -(y + 1/2)$, the phase term cancels out, yielding
\begin{equation} \label{GO_term}
    T_\text{GO} = \sqrt{1 + \frac{1}{y}} \left(1 + \frac{i}{8 \omega y^3} \right) = \sqrt{\mu} \left(1+ \frac{i}{8 \omega y^3} \right) 
\end{equation}
where $\mu = 1 + 1/y$ for the SIS model, and $F(\omega, y)_{\mathrm{GO}} = \sqrt{\mu}$ in the GO limit. 

Thus, the first term in the GO limit (Equation~\ref{GO_term}) clearly corresponds to the classical GO limit. The second term, which scales as $\propto \omega^{-1}$, represents the beyond-GO (bGO) correction to the image, including higher-order expansions of the lensing potential at the image location \citep{2023PhRvD.108d3527T}. The explicit expression for the bGO corrections is provided in Appendix~\ref{bGO}. These corrections are not significant here, as they do not introduce position-dependent phase delays relative to the classical GO term and thus do not induce interference.

\begin{figure*}
    \centering
    \includegraphics[width=\linewidth]{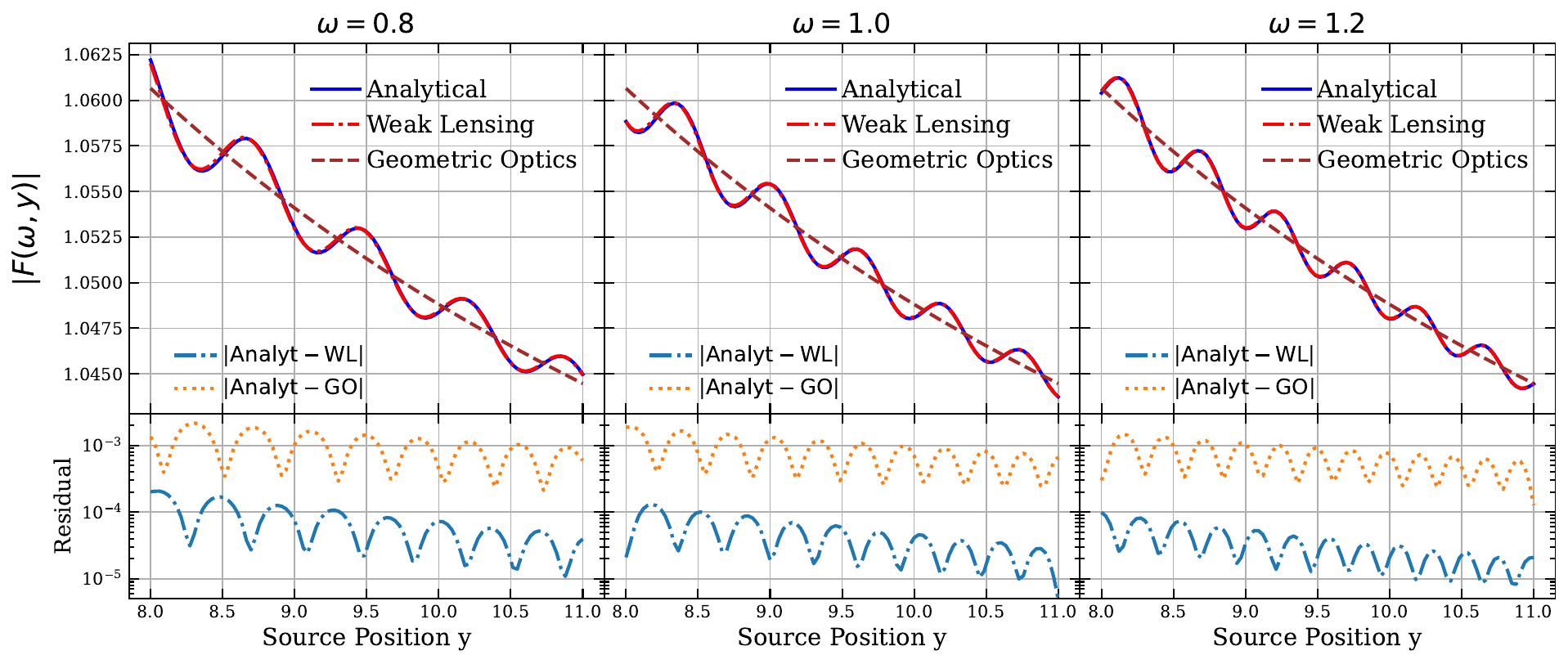}
    \caption{Comparison between the weak lensing approximation and the full analytical results for the SIS lens model. Top: Solid blue lines show the amplification factor magnitude, $|F(\omega, y)|$, as a function of the source position $y$, for $\omega = 0.8$, 1.0, and 1.2 (left, middle, right), computed from the full analytical solution. Dash-dotted red lines represent the weak lensing approximation, and brown dashed lines indicate the GO limit, which is independent of $\omega$. As expected, $|F(\omega, y)|$ approaches the GO limit at large $\omega$ and $y$. Bottom: Dash-dotted blue lines show the absolute difference between the analytical calculation and the weak lensing approximation ($|\text{Analyt} - \text{WL}|$), representing the approximation error. For comparison, dotted orange lines show the absolute difference between the analytical result and the GO limit ($|\text{Analyt} - \text{GO}|$), characterizing the WO term amplitude. The weak lensing approximation shows excellent agreement with the analytical results, with errors consistently below 5\% of the WO term amplitude for $y > 10$.}
    \label{fig:SIS_all}
\end{figure*}

Returning to the second term $T_\text{WO}$, it is associated with the WO effect, 
\begin{equation}
\begin{aligned}
    & T_\text{WO} = e^{-i \omega \phi_m(y)} \sum_{n=0}^{\infty} \frac{\Gamma\left(1+\frac{n}{2}\right)}{n!} \left({- i2\omega }\right)^{n / 2} \\
    & \qquad \times \frac{e^{\frac{i}{2} \omega  y^2} {(\frac{i}{2} \omega  y^2)}^{-\frac{n}{2} - 1}}{\Gamma(-\frac{n}{2})} \left( \sum_{k=0}^{\infty} \frac{\left((1+\frac{n}{2})_k\right)^2}{k!}(\frac{i}{2} \omega  y^2)^{-k}\right) \\
    & = - \frac{2 i}{\omega y^2} e^{i \omega \left( \frac{1}{2}y^2 + y + 1/2 \right)} \sum_{n=0}^{\infty} \left[ \frac{\Gamma\left(1+\frac{n}{2}\right)}{\Gamma\left(1+ n\right) \Gamma(-\frac{n}{2})} \left( \frac{- 2 i }{y}\right)^{n} \right. \\
    & \qquad \left. \times \left( \sum_{k=0}^{\infty} \frac{\left((1+\frac{n}{2})_k\right)^2}{k!}(\frac{i}{2} \omega  y^2)^{-k}\right) \right].
\end{aligned}
\end{equation}
Since $\omega y^2 / 2 \gg 1$, we focus only on the $k=0$ term. An important feature is that when $n = 2m$, with $m \in \mathbb{N}$, $\Gamma\left( -\frac{n}{2} \right) = \Gamma(-m)$ diverges, rendering the term zero. Thus, only odd $n$ contribute, and in particular, the $n=0$ term vanishes. The summation can therefore be further simplified as
\begin{equation}
\begin{aligned}
    &  \sum_{n=0}^{\infty} \frac{\Gamma\left(1+\frac{n}{2}\right)}{\Gamma\left(1+ n\right) \Gamma(-\frac{n}{2})} \left( \frac{-2 i }{y}\right)^{n} \\
    & = - \frac{1}{2} \sum_{m=0}^{\infty} \frac{(2m + 1)  \Gamma\left(1+m+\frac{1}{2}\right)}{\Gamma\left(2+ 2m\right) \Gamma(\frac{1}{2}- m)} \left( \frac{-2 i }{y}\right)^{2m+1}  \\ 
    & = - \frac{1}{2} \sum_{m=0}^{\infty} \frac{ (2m +2)!}{4^{m+1} (-4)^m (1+m)! m!} \left( \frac{-2 i }{y}\right)^{2m+1} \\
    & = - \frac{1}{2 i} \sum_{m=0}^{\infty} \frac{(2m+1)!!}{2m!!} \left( \frac{1}{y}\right)^{2m+1}.
    \end{aligned}
\end{equation}
Since $y \gg y_{\mathrm{cr}} = 1$ by the weak lensing condition for the SIS model, it is sufficient to retain only the $m=0$ term. Thus, the amplification factor for the SIS model under the weak lensing approximation becomes
\begin{equation}
\begin{aligned}
    F(\omega, y) & = \sqrt{1 + \frac{1}{y}} \left(1 + \frac{i}{8 \omega y^3} \right) + \frac{1}{\omega y^3} e^{i \omega \left( \frac{1}{2}y^2 + y + 1/2 \right)} \\
    & \qquad \qquad + \frac{1}{(\omega y^2)} \mathcal{O}(y^{-3}).    \\
    & = \sqrt{\mu}\left(1+\frac{i \Delta_{(1)}}{\omega}\right) + \frac{1}{\omega y^3} e^{i \omega \left( \frac{1}{2}y^2 - \phi_\mathrm{m} \right)} + \ldots\\
    & = T_\mathrm{GO} + T_\mathrm{WO}
\end{aligned}
\end{equation}
Compared to the bGO correction, $T_\mathrm{WO}$ contains position-dependent phase delays relative to the first two terms. This phase variation leads to interference and, in a sense, more characteristically represents the features of WO lensing.

To validate this approximation, we compare it to the full analytical solution expressed in terms of Fresnel integrals (Appendix~\ref{Fresnel}), which is equivalent to Equation~\ref{Eq9} without approximations (Figure~\ref{fig:SIS_all}). The approximation agrees very well with the full solution at large $\omega y^2 / 2$. Specifically, for $y > 10$ and $\omega \sim 1$, the error---defined as the absolute difference ``$|\mathrm{Analyt} - \mathrm{WL}|$"---remains below $5\%$ of the WO features, quantified by the absolute deviation between the analytical WO solution and the GO limit ($|\mathrm{Analyt} - \mathrm{GO}|$). This demonstrates that the weak lensing approximation provides sufficient accuracy for astrophysical applications of WO lensing, as long as the weak lensing conditions are satisfied.

\subsection{General Lenses} \label{Sec_GL}
The SIS model features the simplest lensing potential, $\psi(x) = x$. For general lenses or realistic dark matter distributions, such as the NFW profile, the lensing potential can be significantly more complex. However, as shown by \citet{2004A&A...423..787T}, interference between multiple images gives rise to the oscillatory patterns observed in the $\omega$–$y$ plane.

In the weak lensing regime, even a single observable image can exhibit interference due to an ``imaginary'' image near the lens center \citep{2021MNRAS.507.5390J,2023MNRAS.525.2107J}, which produces the $T_\mathrm{WO}$ term in $F(\omega, y)$ for the SIS model. The WO term is therefore primarily sensitive to the central structure of the lens, whereas the vicinity of the real image governs the GO limit and bGO correction. When $\omega y^2 \gg 1$, the relative phase between the real and imaginary images is dominated by the geometric contribution ($\sim \omega y^2$) and is only weakly dependent on the detailed lens potential \citep{2021MNRAS.507.5390J}. This indicates that the weak lensing approximation can be broadly applicable across different lens models.

To obtain the mathematical prescription of the weak lensing approximation, we have to look into the full integral expression of the amplification factor for general symmetric lenses (Equation~\ref{general_sym}). In the limit $\omega \gg 1$, $F(\omega, y)$ approaches the GO limit and interference is suppressed. Interference should also diminish when $y \gg 1$, as the signal is largely unperturbed by a distant lens. This suppression should be associated with the large-argument limit of the Bessel function in the integrand of amplification factor integral (Equation~\ref{general_sym}), the only term in which $y$ appears apart from the source-dependent constant phase factor $\exp(i \omega \phi_m(y))$.
 
Therefore, it is reasonable to hypothesize that the interference term primarily arises from regions where the argument of the Bessel function is small. Noting that the integral of $x J_0(x)$ yields $x J_1(x)$, and that the asymptotic expansion of Bessel functions becomes valid for $x \lesssim \sqrt{\alpha + 1}$, we focus on the region where $\omega x y$ is small, with $\alpha = 1$. Motivated by this, and noting that all lens information is encoded in the lensing potential, we propose that the potential within the range $0$ to $\sqrt{2}/(\omega y)$ provides the dominant contribution to the WO term.

For a rough but effective approximation, we model the lensing potential as a straight line between $\psi(0)$ and $\psi(\sqrt{2}/(\omega y))$, i.e.,
\begin{equation}
   \psi(x) \approx \left[ \psi\left( \frac{\sqrt{2}}{\omega y} \right) - \psi(0) \right] \frac{\omega y}{\sqrt{2}} x. 
\end{equation}
The resulting approximated potential resembles that of the SIS model, but with a normalization factor that depends on both the impact parameter $y$ and the dimensionless frequency $\omega$. We will present a more detailed physical description towards the end of Section~\ref{discussion}.

The WO interference term for general lenses can thus be approximated as
\begin{equation}
     \mathrm{WO} \sim f(\omega y) \left[ \frac{1}{\omega y^3} e^{i \omega \left( \frac{1}{2}y^2 + y + 1/2 \right)} + \frac{1}{\omega y^3}\mathcal{O}(y^{-2}) \right],  
\end{equation}
where the normalization function $f(\omega y)$ is defined as
\begin{equation} \label{f_def}
    f(\omega y) = \left( \psi\left(\frac{\sqrt{2}}{\omega y}\right) - \psi(0) \right) \frac{\omega y}{\sqrt{2}}.
\end{equation}
The full amplification factor $F(\omega, y)$ remains composed of two parts:
\begin{equation} \label{general_app}
\begin{aligned}
    F&(\omega, y)  = T_\mathrm{GO} + T_\mathrm{WO} \\
    & = \sqrt{\mu}\left(1+\frac{i \Delta_{(1)}}{\omega}\right) + \frac{f(\omega y)}{\omega y^3} e^{i \omega \left( \frac{1}{2}y^2 - \phi_\mathrm{m} \right)} + \ldots
\end{aligned}
\end{equation}
where the expression for $\Delta_{(1)}$ is provided in Appendix~\ref{bGO}. 

This approximation is quite general but requires the lensing potential of the specific model to remain finite at the origin, ensuring the well-definedness of Equation~\ref{f_def}. For instance, the point-mass lens has a potential $\psi \sim \ln x$, which diverges at the center and is therefore incompatible with the weak lensing approximation. It also produces two images even for large $y$, violating the single-image condition. Although Equation~\ref{f_def} resembles a Taylor expansion at the origin, the method does not require $\psi(x)$ to be smooth there. For example, the second derivative of the NFW lensing potential diverges at the origin, yet the prescription still applies; we will validate the accuracy of this approximation using the NFW model as a case study.

\subsection{NFW} \label{Sec_NFW}

\begin{figure*}
    \centering
    \includegraphics[width=\linewidth]{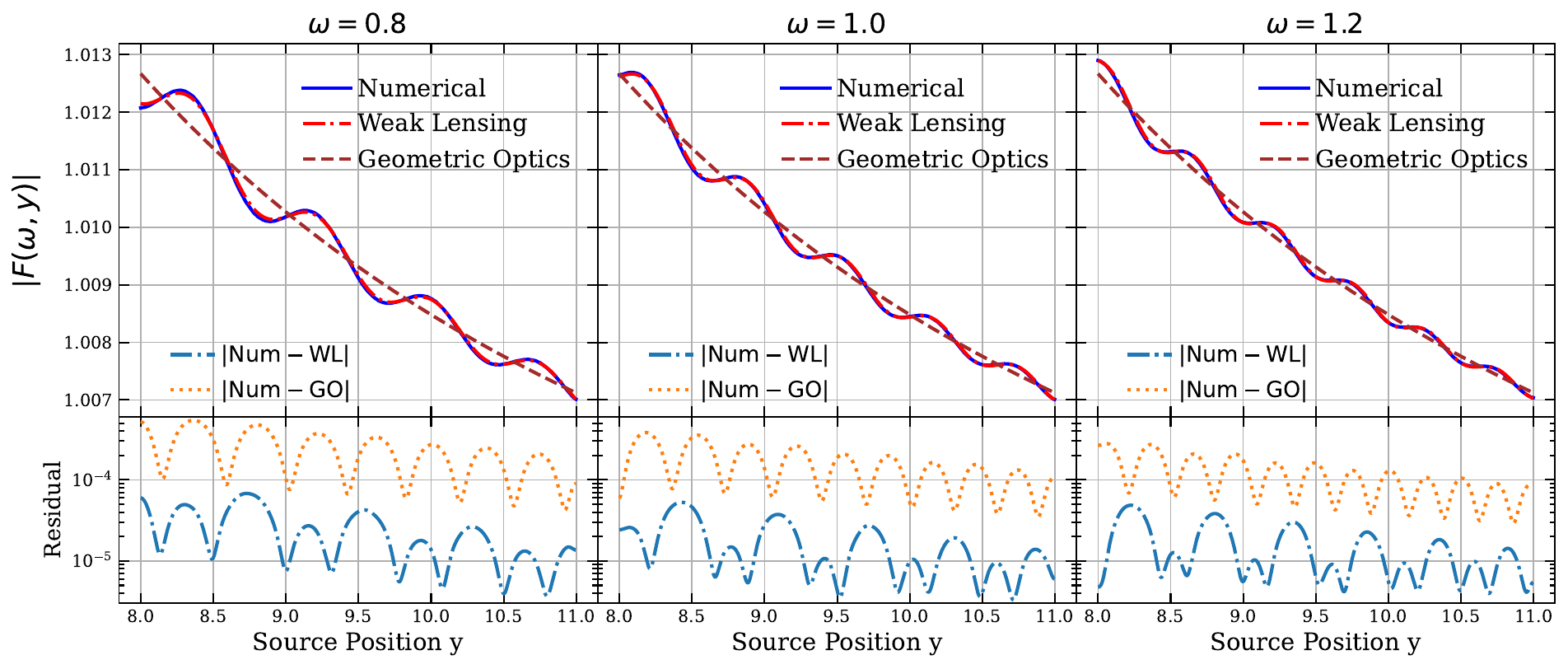}
    \caption{Comparison between the weak lensing approximation and numerical results for the NFW lens model with $\kappa_s = 0.5$. Top: Similar to Figure~\ref{fig:SIS_all}, but the solid blue lines show the amplification factor magnitude, $|F(\omega, y)|$, from numerical calculations performed with \texttt{GRAVELAMPS} \citep{2022ApJ...935...68W} for $\omega = 0.8$, 1.0, and 1.2 (left, middle, right). Bottom: Same as the top panel, but all analytical results are replaced with numerical calculations from \texttt{GRAVELAMPS}, which serve as the benchmark for $|F(\omega, y)|$. For the NFW model, the WO features ($|\text{Num} - \text{GO}|$) at a given source position $y$ are less pronounced compared to the SIS model, reflecting its lower compactness. The weak lensing approximation remains accurate, with errors ($|\text{Num} - \text{WL}|$) limited to a few percent of the WO term amplitude for $y > 10$.}
    \label{fig:NFW_all}
    \vspace{0.5cm} 
\end{figure*}

In the standard concordance cosmological model \citep{2020A&A...641A...1P}, dark matter is assumed to be cold and collisionless. Consistent with these properties, the NFW profile \citep{1997ApJ...490..493N} is widely used to describe the density distribution of cold dark matter halos through a singular, universal scaling function. The density profile is given by
\begin{equation}
    \rho_{\mathrm{NFW}}(r)=\frac{\rho_s}{\frac{r}{r_s}\left(1+\frac{r}{r_s}\right)^2},
\end{equation}
where $\rho_s$ is the characteristic density and $r_s$ is the scale radius. The scale $r_s$ also serves as the normalization constant $\xi_0$ for this profile. Due to the increased complexity of the NFW profile, the lensing potential takes a more complicated form \citep{2003MNRAS.340..105M, 2002A&A...390..821G}:
\begin{equation} \label{nfw_psi}
\psi(x)=2\kappa_s
\begin{cases}
{\left(\ln \frac{x}{2}\right)^2-\left(\operatorname{arctanh} \sqrt{1-x^2}\right)^2} & \text { for } \mathrm{x} < 1 \\ {\left(\ln \frac{x}{2}\right)^2+\left(\arctan \sqrt{x^2-1}\right)^2} & \text { for } \mathrm{x} \geqslant 1
\end{cases}
\end{equation}
where $\kappa_s$ is the dimensionless surface density, defined as $\kappa_s = 4 \pi G \rho_s (D_{\mathrm{L}} D_{\mathrm{LS}} / D_{\mathrm{S}}) r_s/c^2$.

In the WO regime, the $F(\omega,y)$ must be computed numerically from Equation~\ref{general_sym}. Assuming $y \gg 1$\footnote{For the NFW model, $y_{\mathrm{cr}}$ depends on $\kappa_s$. For $\kappa_s = 0.5$, $y_{\mathrm{cr}} \sim 0.17$, while for $\kappa_s = 5$, $y_{\mathrm{cr}} \sim 5.18$. For generality, we adopt $y \gg 1$ as the single-image condition.} ensures the presence of only a single image in the GO limit, where $F(\omega) = \sqrt{\mu}$ and $\mu$ can be approximated as
\begin{equation} 
    \mu \sim 1 + 2\kappa(y) \sim 1 + \frac{4\kappa_s}{y^2-1}[1-\frac{\arctan \sqrt{y^2-1}}{\sqrt{y^2-1}}]
\end{equation}
using the weak lensing limit relations $\mu \sim 1 + 2 \kappa$ and $x_{\mathrm{m}} \sim y$. These approximations also apply to the bGO term\footnote{We verify that the bGO correction is negligible compared to the classical GO and WO terms in the weak lensing limit for the NFW profile and can be safely ignored when approximating $F(\omega, y)$.} $\Delta_{(1)}$, and to the phase at the minimum time delay, where $\phi_{\mathrm{m}} \sim -\psi(y)$.

We approximate the amplification factor $F(\omega, y)$ using Equation~\ref{general_app} and compare the results to numerical calculations performed with the open-source package \texttt{GRAVELAMPS} \citep{2022ApJ...935...68W}, which efficiently and accurately evaluates the amplification factor integral (Equation~\ref{general_sym}) for the NFW lensing potential.

Figure~\ref{fig:NFW_all} shows the comparison for $\omega = 0.8,1.0,1.2$ under the NFW model. The setup is similar to that in Figure~\ref{fig:SIS_all}. As expected, both the GO and WO effects are less prominent for the NFW model compared to the SIS case, reflecting the lower compactness of NFW halos. The weak lensing approximation agrees with the numerical results when $\omega y^2 / 2$ is large. Specifically, for $y > 10$ and $\omega \sim 1$, the error---defined as the absolute difference $|\mathrm{Num - WL}|$---is only a few percent of the WO term amplitude, quantified by the absolute deviation between the numerical results and the GO limits ($|\mathrm{Num - GO}|$). Again, the weak lensing approximation remains sufficiently accurate for astrophysical applications of WO lensing, provided that the weak lensing condition is satisfied. Its successful application to the NFW model further supports the validity of the approximation.

\section{Frequency Asymptotics} \label{freq_asym}
\subsection{High \texorpdfstring{$\omega$}{omega} Limit} \label{high_w}
In our approximation, no explicit condition is imposed on the frequency. Therefore, in the high-frequency limit, the result should recover the expression obtained from the standard high-frequency expansion, provided the weak lensing conditions are satisfied. Specifically, as $\omega \to \infty$, the second weak lensing condition, $\omega y^2 / 2 \gg 1$, is automatically satisfied. However, the first condition, $y \gg 1$, is still required to ensure that only a single image is present.

The recovery of the classical GO limit is straightforward, as both the bGO correction and WO terms scale as $\propto \omega^{-1}$ and vanish at the limit $\omega \to \infty$. The next step is to recover the Quasi-Geometrical Optics (QGO) approximation, which includes the leading-order correction beyond the GO limit and captures interference effects.

For the SIS model with $y > 1$, the QGO approximation is given by \citet{2004A&A...423..787T}:
\begin{equation}
F(\omega, y) \simeq \sqrt{1 + \frac{1}{y}} \left( 1 + \frac{i}{8 \omega y (y + 1)^2} \right) + \frac{e^{i \omega \left[ y^2 / 2 - \phi_{\mathrm{m}}(y) \right]}}{\omega (y^2 - 1)^{3/2}},
\end{equation}
where $\phi_{\mathrm{m}}(y) = -(y + 1/2)$, as defined earlier. This approximation is valid in the high-frequency regime and also holds in the strong lensing regime, although the explicit expression is not provided here. In the weak lensing limit where $y$ is large, we have $y \gg 1$, leading to $(y + 1) \simeq y$ and $(y^2 - 1) \simeq y^2$, in which case the two approximations converge.

For the NFW model, the convergence is more subtle. The diffraction term (WO term) in the QGO approximation is given by \citet{2004A&A...423..787T}:
\begin{equation}
T_\mathrm{WO, QGO} = \frac{4 \kappa_s}{\left( \omega y^2 \right)^2} e^{i \omega \left[ y^2 / 2 - \phi_{\mathrm{m}}(y) \right]},
\end{equation}
where $\kappa_s$ and $\phi_{\mathrm{m}}(y)$ are defined as before. Notice that the phase of the WO term is the same in both approximations, but the magnitudes differ: $4 \kappa_s / (\omega y^2)^2$ in the QGO approximation, and $f(\omega y) / (\omega y^3)$ in the weak lensing approximation. If $f(\omega y) \sim 4 \kappa_s / (\omega y)$ in the relevant limit, the two approximations converge.

\begin{figure}
    \centering
    \includegraphics[width=\linewidth]{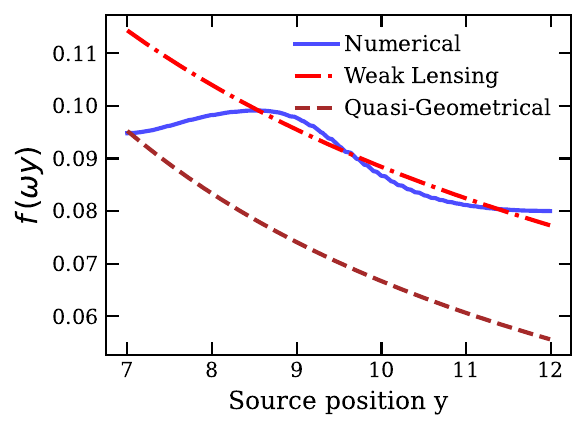}
    \caption{Comparison between the QGO and weak lensing approximations for the normalization function $f(\omega y)$ of the WO term for the NFW model with $\kappa_s = 0.5$ and $\omega = 3$. The red dash-dotted line shows the WO term amplitude from the weak lensing approximation, while the brown dashed line shows the QGO approximation. For comparison, the WO term extracted from numerical calculations using \texttt{GRAVELAMPS} is shown as a blue solid line. The numerical results indicate that the QGO approximation performs better at small $y$, while the weak lensing approximation is more accurate at larger $y$, consistent with theoretical expectations.}
    \label{fig:QGO_WL}
\end{figure}

Figure~\ref{fig:QGO_WL} shows the normalized magnitude (after division by $\omega y^3$) of the WO term under the QGO and weak lensing approximations as a function of source position. The magnitude inferred from numerical calculations using \texttt{GRAVELAMPS} is also shown. As expected, the QGO approximation performs better at small $y$, while the weak lensing approximation becomes more accurate at larger $y$. For sufficiently large $\omega$ and $y$, both approximations and the numerical results converge. This convergence arises because, in the high-frequency and large-$y$ limit, the ``imaginary image" in both approximations is sensitive only to the mass distribution near the center of the lens \citep{2004A&A...423..787T}.

The high-frequency limit is relevant when a GW passes near a massive structure, such as a galactic dark matter halo with mass $M \sim 10^{12} M_\odot$ (see Equation~\ref{w_limit} for the SIS case). However, the WO effect is suppressed at large $\omega$ and $y$, and the number density of massive halos is significantly lower than that of less massive ones. According to \citet{2023PhRvD.108l3543C}, most detectable weak lensing WO events for SIS lenses occur outside the high-frequency regime. Future work should extend these studies to more realistic profiles, such as NFW halos.

\subsection{Low \texorpdfstring{$\omega$}{omega} Limit} \label{low_w}
If a GW passes near low-mass structures, such as subhalos or isolated halos with virial mass $M_{\mathrm{vir}} \sim 10^6 \, M_\odot$, the lensing falls into the low-frequency regime. For SIS lenses, the dimensionless frequency $\omega$ is related to the virial mass by \citep{2023PhRvD.108j3532S}:
\begin{equation} \label{w_limit}
\begin{aligned}
    \omega \sim 6 \times 10^{-4} \left(\frac{M_\mathrm{vir}}{10^6 \, M_{\odot}}\right)^{4/3}\left(\frac{f}{10 \, \mathrm{mHz}}\right) \times \\
 \times (1 + z_L)^2\left(\frac{d_{\mathrm{eff}}}{1 \,\mathrm{Gpc}}\right) \left(\frac{H(z_L)}{H_0}\right)^{4/3}
\end{aligned}
\end{equation}
where $f$ is the GW frequency, $d_{\mathrm{eff}}$ is defined in Equation~\ref{M_Lz}, and $H(z_L)$ is the Hubble parameter at redshift $z_L$. In the low-$\omega$ limit, $F(\omega)$ approaches unity, as the GW wavelength exceeds the lens' characteristic scale and the wave is essentially unperturbed \citep{2023PhRvD.108d3527T}. 

In the weak lensing approximation, this behavior is naturally recovered, as the approximation is valid only when $\omega y^2 / 2 \gg 1$. For example, keeping $\omega y^2 / 2$ large and fixed, a small $\omega \ll 1$ requires $y \gg 1$ to maintain the condition. According to Equation~\ref{general_app}, and noting that $\mu \propto 1/y^2$ and $\Delta_{(1)} \propto 1/y^4$ for the NFW model, both the bGO\footnote{In \citet{2004A&A...423..787T}, the bGO expansion is discussed in the high-frequency limit. However, when the variation of the lensing potential is small compared to the dimensionless frequency, as is the case in the weak lensing limit, the bGO expansion remains valid as the expansion condition is still satisfied.} and WO terms vanish as $\omega y^3 \gg 1$, even at fixed $\omega y^2$. In contrast, the QGO approximation yields a WO term with amplitude $4 \kappa_s / (\omega y^2)^2$, which remains finite for fixed $\omega y^2$ even as $\omega \to 0$. Systems with $\omega y^2 / 2 \lesssim 1$ are not captured by our weak lensing approximation and may instead be treated with alternative approximations such as \citet{PhysRevD.104.063001}.

\begin{figure}
    \centering
    \includegraphics[width=\linewidth]{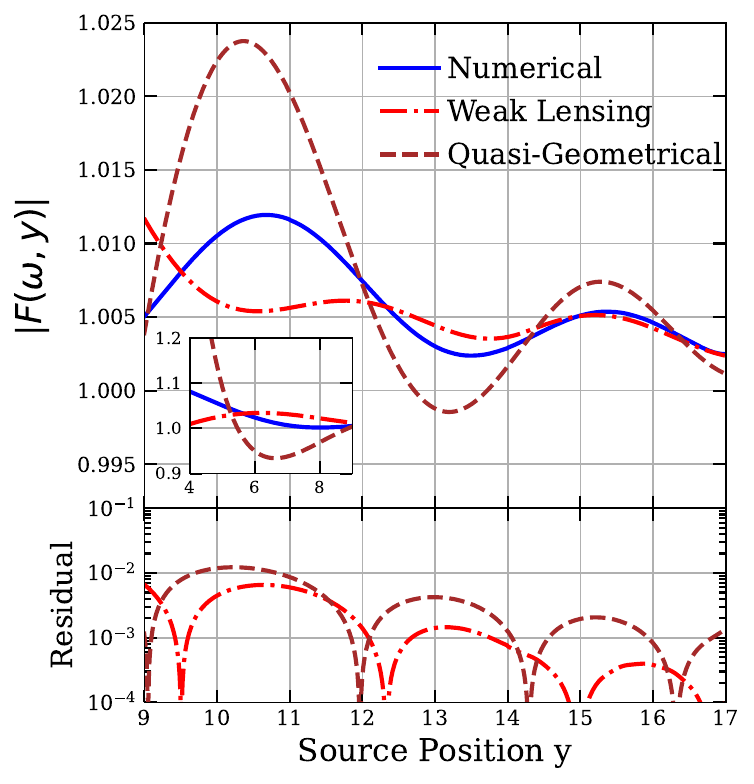}
    \caption{Comparison between the weak lensing approximation, QGO approximation, and numerical results for the NFW lens model with $\kappa_s = 0.5$ and $\omega = 0.1$. The layout of the top and bottom panels follows Figure~\ref{fig:NFW_all}, but here the brown dashed line shows the amplification factor magnitude, $|F(\omega, y)|$, from the QGO approximation \citep{2004A&A...423..787T}. An inset in the top panel highlights the range $4 \leq y \leq 9$. The weak lensing approximation outperforms the QGO approximation at low $\omega$, particularly at small $y$, and provides an accurate description once $\omega y^2/2 \gtrsim 10$.}
    \label{fig:NFW_0.1}
\end{figure}

The weak lensing approximation also performs well at low $\omega$, before reaching the fully unperturbed limit. Figure~\ref{fig:NFW_0.1} shows the comparison for $\omega = 0.1$ under the NFW model, with both the QGO and weak lensing approximations compared to the numerical results. The QGO approximation becomes inaccurate at low $\omega$, diverging for $y < 5$. In contrast, the weak lensing approximation remains stable and agrees well with the numerical results once $\omega y^2 / 2 \sim 10$ is reached. This demonstrates the robustness of the weak lensing approximation for lensing by low-mass structures.

\section{Discussion} \label{discussion}
Gravitational lensing in the weak lensing regime is particularly important for realistic GW detections, as most standard sirens at high redshift are expected to be affected by weak lensing. However, the weak lensing signal is dominated by the collective effect of numerous dark matter halos at large impact parameters. Even when neglecting interference between lenses and simply summing individual contributions, obtaining the amplification factor from populations of general lenses remains computationally expensive using numerical methods. For each amplification factor, it requires evaluating complex families of contours in the time domain \citep{2025PhRvD.111j3539V}. 

To address this computational challenge, several analytic approximation schemes have been proposed. \citet{PhysRevD.104.063001} introduced a method to approximate the amplification factor in the weak diffractive lensing regime, applicable in the low-frequency limit where $\omega y^2 \lesssim 1$. \citet{2025PhRvD.111h3541Y} further generalized this approach beyond the strictly diffractive limit, but their formulation still assumes small $\omega$ and neglects the contribution from the central imaginary image, which is one of the most distinctive and physically important features of WO effects in the weak lensing regime and is properly captured by our approximation. On the other hand, \citet{2023PhRvD.108j3532S} use the weak lensing condition to simplify the contours, but their method still involves one-dimensional numerical integration for each approximated contour. In contrast, our analytical weak lensing approximation requires no numerical integrations and is significantly faster.

Another major advantage of our method is that most numerical approaches evaluate the amplification factor in the time domain first, including the approach where a weak lensing approximation is implemented. However, transforming back to the frequency domain is nontrivial. Efficient computation requires Fast Fourier Transform techniques with adaptive resolution, and careful regularization is necessary to avoid artifacts in the frequency-domain results \citep{2023PhRvD.108d3527T, 2023PhRvD.108j3529T, 2025PhRvD.111j3539V}. These challenges become particularly severe at large $y$ or $\omega$. In contrast, our method operates entirely in the frequency domain, avoiding the need for adaptive resolution and notoriously annoying regularizations, and is thus free from these complications.

Although the weak lensing approximation appears simple, it offers valuable physical insights. The approximation of the WO term involves only the lensing potential at $x = 0$ and $x = \sqrt{2}/(\omega y)$, where $x$ is the normalized impact parameter. In other words, only the inner region of the lens contributes significantly to the WO term in the weak lensing limit, and the size of this effective region decreases with increasing $y$ and $\omega$. 

To gain further insight into the weak lensing approximation, the connection between the WO term and the physical properties of the lens can be made explicit by expressing it in terms of the enclosed mass. For symmetric lenses, the deflection angle is
\begin{equation}
|\boldsymbol{\alpha}(x)|=\frac{2}{x}  \int_0^{x} \mathrm{d} x^{\prime} x^{\prime} \kappa\left(x^{\prime}\right) \propto \frac{M_{<x}}{ x} 
\end{equation}
where $\kappa(x) = \Sigma(\xi_0 x)/\Sigma_{\mathrm{cr}}$ is the normalized surface mass density (convergence), and $M_{<x}$ is the mass enclosed within $x$. The lensing potential is obtained by integrating the deflection angle, with the reference point chosen such that $\psi(0) = 0$. In the weak lensing approximation, the potential is treated as linear within a small region, so the normalization function $f(\omega y)$ represents the average derivative of the lensing potential, proportional to the enclosed mass $M_{<\sqrt{2}/(\omega y)}$ divided by its critical scale $\sqrt{2}/(\omega y)$. The WO term then takes the form
\begin{equation}
    T_\text{WO} \propto \frac{M_{<\frac{\sqrt2}{\omega y}}}{y^2} e^{i \omega \left( \frac{1}{2} y^2 - \phi_{\mathrm{m}} \right)}.
\end{equation}
This establishes a direct relation between the enclosed mass and the strength of the WO term. During the inspiral of a compact binary, the GW frequency increases with time, causing the WO term to probe the central mass of the lens on progressively smaller scales---effectively performing a tomographic scan of the lens.

Physically, the weak lensing approximation of the WO term can be described as follows: the frequency and impact parameter of the GW signal determine a critical scale of the lens, within which the mass distribution can be approximated by an SIS-like profile. The normalization of the effective SIS lens is set by the normalized mass enclosed within this critical radius. This approximation agrees well with precise numerical calculations, which are significantly more time-consuming and suffer from numerical instabilities when $\omega$ and $y$ are large.

In the GO limit, weak lensing is primarily sensitive to the local projected density and not to the detailed mass profile of the lens. In the WO regime, the situation changes due to the appearance of an ``imaginary image''\footnote{\citet{2004A&A...423..787T} referred to the ``imaginary image'' as the diffracted image, though they represent the same physical phenomenon.} formed at the lens center by diffraction. Nevertheless, the lensing signal remains largely insensitive to the detailed mass profile, depending mainly on the local projected density---even for the imaginary image that contributes to the WO term. This property is advantageous for model-agnostic applications such as delensing standard sirens, as it reduces dependence on the assumed lens model. However, it poses challenges for efforts to infer the very inner structure of the lens from weakly lensed signals.

\section{Conclusion and Outlook} \label{conclu}
In this paper, we have presented an analytical method to approximate the amplification factor $F(\omega, y)$ under the weak lensing conditions $\omega y^2 / 2 \gg 1$ and $y \gg 1$\footnote{Orders of several tens are considered large in this work. The latter condition can be further relaxed if $y_{\mathrm{cr}} \lesssim 1$.}. The approximation is particularly useful for moderately low $\omega$ with moderately large $y$, making it well-suited for studying WO effects caused by low-mass field halos and subhalos. 

The method is valid for general symmetric lens profiles, including both SIS and NFW models, and is given by
\begin{equation} \label{final_general}
F(\omega, y) \simeq \sqrt{\mu} \left( 1 + \frac{i \Delta_{(1)}}{\omega} \right) + \frac{f(\omega y)}{\omega y^3} e^{i \omega \left( \frac{1}{2} y^2 - \phi_{\mathrm{m}} \right)},
\end{equation}
where the normalization function $f(\omega y)$ is related to the lensing potential $\psi$ by Equation~\ref{f_def}, and $\Delta_{(1)}$ is defined in Appendix~\ref{bGO}.

For the SIS model, the weak lensing approximation reduces to
\begin{equation}
F(\omega, y) = \sqrt{1 + \frac{1}{y}} \left( 1 + \frac{i}{8 \omega y^3} \right) + \frac{1}{\omega y^3} e^{i \omega \left( \frac{1}{2} y^2 + y + \frac{1}{2} \right)},
\end{equation}
while for the NFW model, the functions $f(\omega y)$ and $\Delta_{(1)}$ are more complicated but remain fully analytical. The weak lensing approximation for both SIS and NFW models shows excellent agreement with numerical results obtained using the open-source package \texttt{GRAVELAMPS} \citep{2022ApJ...935...68W}. Additionally, we have demonstrated that the weak lensing approximation correctly recovers both the low- and high-frequency limits.

Due to limitations in numerical calculations, current studies of WO effects on low-frequency GW signals have been largely restricted to simple point-mass and SIS lens models \citep{2022MNRAS.512....1G, 2023PhRvD.108l3543C, 2023PhRvD.107d3029C, 2023PhRvD.108j3532S}. With the analytical weak lensing approximation developed here, it becomes feasible to statistically study more realistic lens profiles, such as the NFW and Einasto models, and to make more robust predictions for future GW observations.

The simple expression proposed in this work opens the possibility for systematic lens reconstruction in the weak lensing regime. By assuming a lens profile for the halos, one can infer both $\omega$ and $y$ by simultaneously fitting the magnitude and phase of the WO term from the detected amplification factor. This cannot be achieved with the GO approximation, which produces only frequency-independent magnification. For multiple lenses, following the additive property of lensing potentials, the collective WO effects can be obtained by simple summation of the individual WO terms.

The weak lensing approximation for WO effects is particularly valuable for probing low-mass halos, where strong lensing is highly unlikely. This approach enables studies of the mass and spatial distributions of halos with negligible baryonic content, providing clean probes to test the nature of dark matter. However, such reconstructions will ultimately be limited by the signal-to-noise ratio, and detection of the amplification factor is subject to potential degeneracies among lens and source parameters \citep{2023PhRvD.108l3543C}.

Accurate modeling of WO effects can also enable recalibration of the intrinsic luminosity of the source, as demonstrated by \citet{2023PhRvD.108j3532S} for the SIS model. This would reduce a major source of uncertainty in using standard sirens as cosmological probes and, importantly, does not require dedicated electromagnetic follow-up observations, unlike traditional so-called ``delensing'' methods \citep{2010MNRAS.404..858S, 2023MNRAS.522.4059W}.

Additionally, our method can be applied to nested symmetric lenses, such as a chameleon profile for baryonic matter embedded within a dark matter halo described by an NFW profile \citep{2020A&A...643A.135C}, or an NFW lens with embedded subhalos \citep{2019Galax...7...81Z}. Furthermore, the method may be extended to non-symmetric lens profiles via perturbative techniques, provided the deviation of the lensing potential from axisymmetry is small---a condition thought to be common for galactic dark matter halos based on observational evidence \citep{2006glsw.conf.....M}.

\section*{Acknowledgments}
O.A.H. acknowledges support by grants from the Research Grants Council of Hong Kong (Project No. CUHK 14304622, 14307923, and 14307724), the start-up grant from the Chinese University of Hong Kong, and the Direct Grant for Research from the Research Committee of The Chinese University of Hong Kong. M. H. acknowledges support of the Science and Technology Facilities Council (Grant Ref. ST/V005634/1). Q. L. N. acknowledges support of the Institute for Cosmic Ray Research International University Research Program. This research has made use of NASA's Astrophysics Data System Bibliographic Services.

%



\appendix
\section{beyond GO correction} \label{bGO}
The first-order correction to the GO limit is given by
\begin{equation}
F(\omega) \simeq \sqrt{|\mu|}\left(1+\frac{i \Delta_{(1)}}{\omega}\right)
\end{equation}
where $\mu$ is the magnification of the image. For symmetric lenses, $\Delta_{(1)}$ is given by \citep{2023PhRvD.108d3527T, 2023PhRvD.108j3532S}:
\begin{equation}
|\mu|^{-1} \equiv 4 a b, \, \, \, \Delta_{(1)} \equiv \frac{1}{16}\left[\frac{\psi^{(4)}}{2 a^2}+\frac{5}{12 a^3}\left(\psi^{(3)}\right)^2+\frac{\psi^{(3)}}{a^2 x}+\frac{a-b}{a b x^2}\right] 
\end{equation}
where $a \equiv (1 - \psi'') / 2$, $b \equiv (1 - \psi' / x) / 2$, and $\psi^{(n)} \equiv \mathrm{d}^n \psi / \mathrm{d}x^n$. All quantities are evaluated at the image position $x = x_{\mathrm{m}}$.  See \cite{2023PhRvD.108j3532S} for higher orders of the bGO corrections.

\section{SIS Amplification factor with Fresnel Integral Representation} \label{Fresnel}
While a closed-form expression for $F(\omega, y)$ for the SIS model is not available, it can be reduced to a single angular integral following \citet{2023PhRvD.108j3532S}. The final result is
\begin{equation}
F(w)=e^{i w\left(y^2 / 2-\phi_m\right)}[1 +\int_0^\pi \mathrm{d} \theta \alpha f(-\alpha)  \left.-i \int_0^\pi \mathrm{d} \theta \alpha g(-\alpha)\right]
\end{equation}
where
\begin{equation}
    \alpha(\theta) \equiv \sqrt{\frac{w}{\pi}}\left(\psi_0+y \cos \theta\right)
\end{equation}
and $f(z)$ and $g(z)$ are auxiliary functions defined in terms of the Fresnel sine $S(z)$ and cosine $C(z)$ functions:
\begin{equation}
\begin{aligned}
f(z) \equiv & \left(\frac{1}{2}-S(z)\right) \cos \left(\frac{\pi}{2} z^2\right)
-\left(\frac{1}{2}-C(z)\right) \sin \left(\frac{\pi}{2} z^2\right) \\
g(z) \equiv & \left(\frac{1}{2}-C(z)\right) \cos \left(\frac{\pi}{2} z^2\right) +\left(\frac{1}{2}-S(z)\right) \sin \left(\frac{\pi}{2} z^2\right).
\end{aligned}  
\end{equation}

\bibliography{Zhaofeng_lib,references_otto_additional}{}
\bibliographystyle{aasjournal}



\end{document}